# Photonic RF and microwave arbitrary waveform generator based on a soliton crystal 49GHz Kerr micro-comb

Mengxi Tan, Xingyuan Xu, Andreas Boes, Bill Corcoran, Jiayang Wu, *Member, IEEE,* Thach G. Nguyen, Sai T. Chu, Brent E. Little, Roberto Morandotti, *Senior Member, IEEE, Fellow OSA,* Arnan Mitchell, *Senior Member, IEEE, and* David J. Moss, *Fellow, IEEE, Fellow OSA*

*Abstract*— We report a photonic-based radio frequency (RF) arbitrary waveform generator (AWG) using a soliton crystal micro-comb source with a free spectral range (FSR) of 48.9 GHz. The comb source provides over 80 wavelengths, or channels, that we use to successfully achieve arbitrary waveform shapes including square waveforms with a tunable duty ratio ranging from 10% to 90%, sawtooth waveforms with a tunable slope ratio of 0.2 to 1, and a symmetric concave quadratic chirp waveform with an instantaneous frequency of sub GHz. We achieve good agreement between theory and experiment, validating the effectiveness of this approach towards realizing high-performance, broad bandwidth, nearly user-defined RF waveform generation.

## I. INTRODUCTION

Radio frequency arbitrary waveform generators (AWGs) are an important category of signal sources that can generate arbitrary user-defined waveforms. They have attracted significant attention in many applications ranging from wireless communications, to radar, measurement systems, and others [1-3]. Most approaches to RF AWGs are based on electronic technologies and while these are very mature with advanced instrumentation available commercially, they are bulky, very expensive, and subject to limitations in speed and linearity due largely to the required use of digital-to-analogue converters [4].

Photonic approaches [5-7], on the other hand, offer very high bandwidths and low phase noise that are difficult to obtain purely electronically [4]. There are a wide variety of approaches to realizing photonic RF arbitrary waveform generation [4, 8-15], such as spatial-to-temporal mapping [8, 9], wavelength-to-time mapping [4, 10], and Fourier synthesis [11] based on line-by-line control of optical frequency combs [12, 13]. However, while offering many advantages, these approaches face challenges of one form or another. For the spatial-to-time mapping and wavelength-to-time mapping AWGs, the synthesized waveforms are generally single-shot pulses, which are not suitable for RF applications that require continuous waveforms. For the Fourier synthesis approach, the synthesized signal bandwidth is subject to the resolution of the line-by-line spectral shaping, thus is unable to reach low-frequency RF bands.

Integrated optical Kerr frequency comb sources, or 'micro-combs' [16-22], have come into focus as a fundamentally new and powerful tool due to their ability to provide highly coherent multiple wavelength sources for RF applications [23-30]. They can greatly increase the capacity of communications systems and allow the processing of RF spectra for a wide range of advanced signal processing functions [31-35]. Micro-combs have the potential to provide a much larger number of wavelengths, an ultra-large Nyquist bandwidth compared to mode-locked lasers, as well as to offer a greatly reduced footprint and degree of complexity.

RF transversal filtering is a powerful approach to RF signal processing that is particularly well suited to multiwavelength optical sources [24]. For RF transversal filter functions, the number of wavelengths dictates the available number of channels to provide the RF time delays. Thus, with micro-combs, the performance of RF signal processing functions and other systems such as beamforming devices can be greatly enhanced in terms of the quality factor and angular resolution. In particular, the combination of a low FSR (for microcombs) of 50GHz or less, together with the use of soliton crystals, has proven extremely successful for a wide range of RF photonic applications [25-33]. Based on these advantages, a wide range of RF applications have been demonstrated, such as optical true time delays [25], transversal filters [26-28], signal processors [29-31], channelizers [34, 35], and phase-encoded signal generators [33]. Previously, we reported a photonic RF phase-encoded signal generator that achieved a phase encoding rate ranging from 1.98 to 5.95 Gb/s in a compact footprint [22]. In that approach, an RF single-cycle pulse was multicast onto a spectrally flattened micro-comb, with the progressively delayed replicas assembled arbitrarily in time according to the designed binary phase codes.

In this letter, we propose and demonstrate a user-defined RF arbitrary waveform generator based on a soliton crystal micro-comb source. Eighty-one wavelength channels are used, compared with 60 wavelengths for our previous work [33], which significantly enhances the speed and flexibility of our signal generation system. We present the architecture of the arbitrary waveform generator and then the

This work was supported by the Australian Research Council Discovery Projects Program (No. DP150104327). RM acknowledges support by the Natural Sciences and Engineering Research Council of Canada (NSERC) through the Strategic and Discovery Grants Schemes, by the MESI PSR-SIIRI Initiative in Quebec, and by the Canada Research Chair Program. Brent E. Little was supported by the Strategic Priority Research Program of the Chinese Academy of Sciences, Grant No. XDB24030000.

M. Tan, X. Xu, J. Wu, and D. J. Moss are with the Optical Sciences Centre, Swinburne University of Technology, Hawthorn, VIC 3122, Australia. X. Xu is currently with the Electro-Photonics Laboratory, Department of Electrical and Computer Systems Engineering, Monash University, VIC3800, Australia. (Corresponding e-mail: dmoss@swin.edu.au).

A. Boes, T. G. Thach and A. Mitchell are with the School of Engineering, RMIT University, Melbourne, VIC 3001, Australia.

B. Corcoran is with the Department of Electrical and Computer System Engineering, Monash University, Clayton, VIC 3168 Australia.

S. T. Chu is with the Department of Physics, City University of Hong Kong, Tat Chee Avenue, Hong Kong, China.

B. E. Little is with the State Key Laboratory of Transient Optics and Photonics, Xi'an Institute of Optics and Precision Mechanics, Chinese Academy of Science, Xi'an, China.

R. Morandotti is with INRS - Énergie, Matériaux et Télécommunications, 1650 Boulevard Lionel-Boulet, Varennes, Québec, J3X 1S2, Canada, and an adjunct with the Institute of Fundamental and Frontier Sciences, University of Electronic Science and Technology of China, Chengdu 610054, China.

results of experiments generating user-defined waveforms. These include a tunable square waveform with a duty ratio ranging from 10% to 90%, sawtooth waveforms with tunable slope ratios from 0.2 to 1, and symmetric concave quadratic chirp waveforms with an instantaneous frequency reaching down to the sub GHz range. Our successful results stem from the soliton crystal's extremely robust and stable operation and generation, as well as its much higher intrinsic efficiency, all of which are enabled by an integrated CMOS compatible platform. The high performance as well as the good agreement between theory and experiment confirms our approach as being an effective way to implement user-defined RF waveform generation with reduced footprint, lower complexity, and potentially lower cost.

## II. INTEGRATED KERR MICROCOMB GENERATION

The generation of micro-combs is a complex process that generally relies on a high nonlinear material refractive index, low linear and nonlinear loss, as well as engineered anomalous dispersion [38-44]. Diverse platforms have been developed for micro-comb generation [22], such as silica, magnesium fluoride, silicon nitride, and doped silica glass. The micro-ring resonators (MRRs) used to generate the Kerr optical micro-combs in the experiments reported here were fabricated on a high-index doped silica glass platform using CMOS compatible processes. Due to the ultra-low loss of our platform, the MRR features narrow resonance linewidths, corresponding to quality factors as high as ~1.5 million., with radii of ~592 μm, which corresponds to a very low FSR of ~0.393 nm (~48.9 GHz) [24, 27, 30-34]. For the fabrication process, first, high-index ($n = $~1.7 at 1550 nm) doped silica glass films were deposited using plasma-enhanced chemical vapour deposition, which were then patterned by deep ultraviolet photolithography and reactive ion etching to form waveguides with exceptionally low surface roughness. Finally, silica ($n = $ ~1.44 at 1550 nm) was deposited as an upper cladding. The advantages of our platform for optical micro-comb generation include ultra-low linear loss (~0.06 dB·cm$^{-1}$), a moderate nonlinear parameter (~233 W$^{-1}$·km$^{-1}$) and, in particular, a negligible nonlinear loss up to extremely high intensities (~25 GW·cm$^{-2}$) [45-56]. After packaging the device with fibre pigtails, the through-port insertion loss was as low as 0.5 dB/facet, assisted by on-chip mode converters.

To generate soliton crystal micro-combs, we amplified the pump power up to 30.5 dBm. When the detuning between the pump wavelength and the cold resonance became small enough, such that the intra-cavity power (Fig. 1) reached a threshold value, modulation instability (MI) driven oscillation was initiated. Primary combs (Fig.1 (b) (c)) were thus generated with the spacing determined by the MI gain peak – mainly a function of the intra-cavity power and dispersion. As the detuning was changed further, distinctive 'fingerprint' optical spectra [41, 42] were observed (Fig. 1 (d)). The spectra are similar to what has been reported from spectral interference between tightly packed solitons in the cavity – so-called 'Soliton Crystals' [24, 25]. The second step in the measured intracavity power was observed at the point where their spectra appeared. Nonetheless, the soliton crystal states provided the lowest noise states of all our micro-combs and were used as the basis for a microwave oscillator with low phase-noise [32].

## III. PHOTONIC RF ARBITRARY WAVEFORM GENERATION

Figure 2 and 3 illustrates the operation principle of the photonic RF signal generation. First, an RF Gaussian pulse $f(t)$ with a duration of $\Delta t$ is generated. The flattened comb lines are fed into an intensity modulator which, when driven by the input RF signal, yields replicas of the RF pulse in the optical domain – effectively multi-casting the RF signal onto many wavelengths at once. The modulated signal produced by the intensity modulator is then directed through a spool of dispersive elements, generating a time delay $\Delta t$ between adjacent wavelengths. Next, the wavelengths are programmably shaped and separated using a WaveShaper according to the designed taps weights $g[n]$ (with length $N$). Finally, the weighted replicas are summed upon photodetection using a high-speed balanced photodetector (in our case, with a 43 GHz bandwidth), thus achieving a discrete convolution operation between the RF pulse and shaped microcomb spectrum, described as:

$$(f*g)[n] = \sum_{i=1}^{N} f[n - i \cdot \Delta t] \cdot g[i] \quad (2.1)$$

In order to generate the user-defined RF waveform, the soliton crystal micro-comb was first flattened and modulated with the RF input pulse (Fig. 3 (a)) in order to multicast the RF waveform onto all of the wavelength channels to yield 81 replicas. These were then transmitted through a spool of standard signal mode fibre ($L = 13$ km, $\beta = $ ~17.4 ps/nm/km) to obtain a progressive time delay between adjacent wavelengths of $\Delta t = $ ~89 ps in order to roughly match the input pulse width. Next, the WaveShaper (Finisar 4000s) accurately shaped the comb power according to the designed taps weights, with the shaped comb spectrum shown in Fig. 3 (b). The wavelength channels for positive and negative taps were separately measured by an optical spectrum analyser. The optical power for each comb line closely matched the designed tap weights, verifying the success of our comb shaping procedure. Finally, the delayed replicas were combined and then summed upon photo-detection. Both positive and negative taps were achieved by separating (spatially demultiplexing) the wavelength channels according to the sign of the designed tap weights and then fed into a balanced photodetector (Finisar BPDV2150R). By tailoring the comb lines' power according to the tap weights, arbitrary waveform generation (Fig. 3 (c)) could be achieved. The electrical signal has the same shape as the optical power spectrum, with measured waveforms being normalized to the peak intensity.

To demonstrate the flexibility of our photonic RF signal generation approach, we designed square waveforms (Fig. 4 (a), Fig. 5 (a)) with a tunable duty cycle ratio ranging from 10% to 90%. Similarly, sawtooth waveforms (Fig. 4 (b), Fig. 5 (b)) with a tunable slope ranging from 0.2 to 1 were generated. The received signals were digitally sampled in an 80 GSa/s real-time oscilloscope, with the measured waveforms normalized to the peak intensity. The shaped comb spectra (Fig. 4 (a) and (b)) show that the optical power for each comb line closely matched the designed tap weights, with the same shape as the designed RF waveform, which makes the user-defined generation easy and flexible. We then demonstrated the frequency-modulated waveform, as shown in Fig. 6, for which the sign of the frequency modulation (or 'chirp') can be programmed to sweep from high to low and then from low to high, which is very difficult to achieve with electronic techniques. We compared the experimental results obtained with the corresponding calculated instantaneous frequency of the designed symmetric concave quadratic chirp, both of which are shown in Fig. 6 (c, d).

Compared to electronic means of arbitrary waveform generation, our scheme makes possible RF waveforms with much higher instantaneous bandwidth by simply shortening the time delay and the corresponding optical pulse width. Note that in this work we used a commercial arbitrary waveform generator (Keysight, 65 GSa/s) to generate the pulse for this proof of principle demonstration. This allowed us to investigate the device performance by, for example, readily changing the pulse width to test the capability of our system to span different frequency ranges for the RF waveform. In practice, however, electronic AWGs are not necessary and can easily be replaced with many other readily available approaches that are simpler, easier and cheaper [13].

IV. CONCLUSION

We demonstrate a photonic based method for the generation of RF arbitrary waveforms. The high-quality, versatility, and performance of the measured waveforms matches well with theory. This approach provides simple operation and fast reconfiguration, thus offering significant benefits in the application of photonic based arbitrary waveform generation.

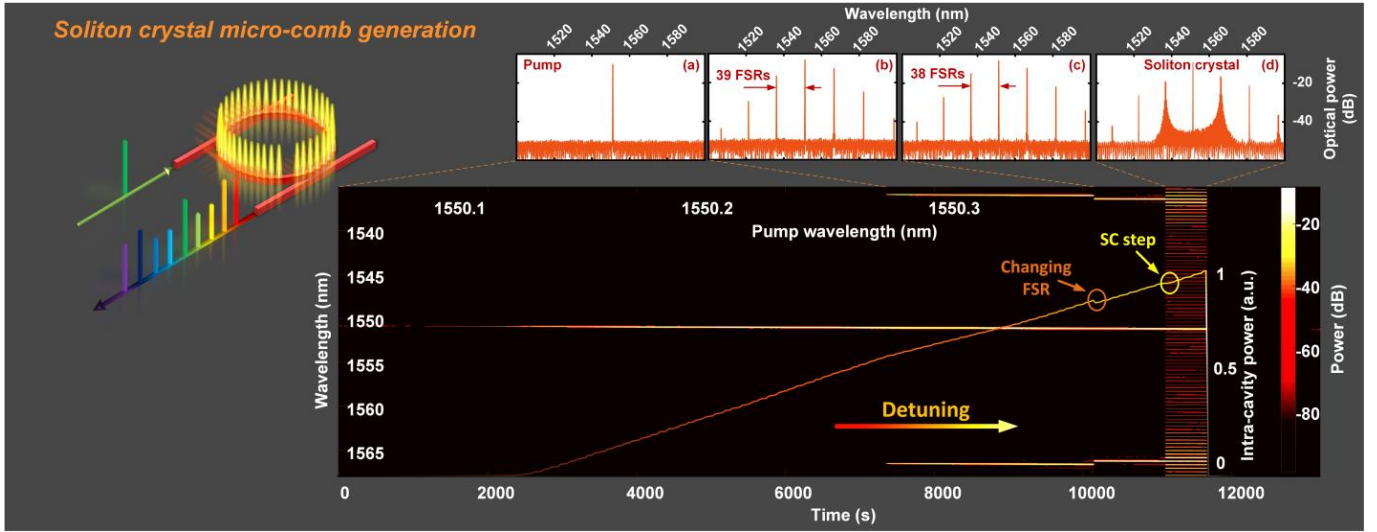

Fig. 1. Conceptual diagram of soliton crystal micro-comb generation. Optical spectra of (a) Pump. (b) Primary comb with a spacing of 39 FSRs. (c) Primary comb with a spacing of 38 FSRs. (d) Soliton crystal micro-comb.

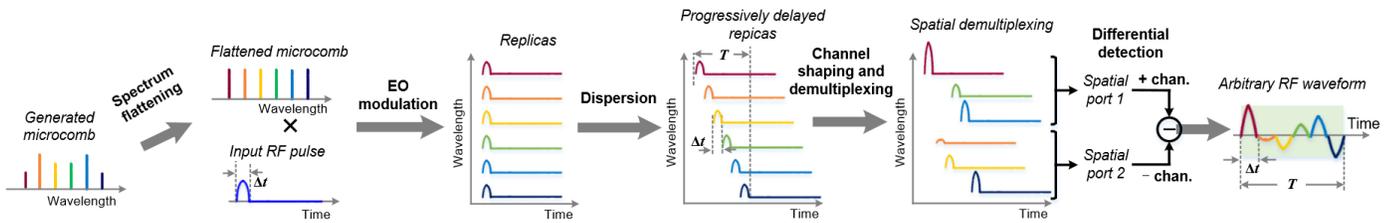

Fig. 2.　Illustration of the operation of the photonic arbitrary waveform generator.

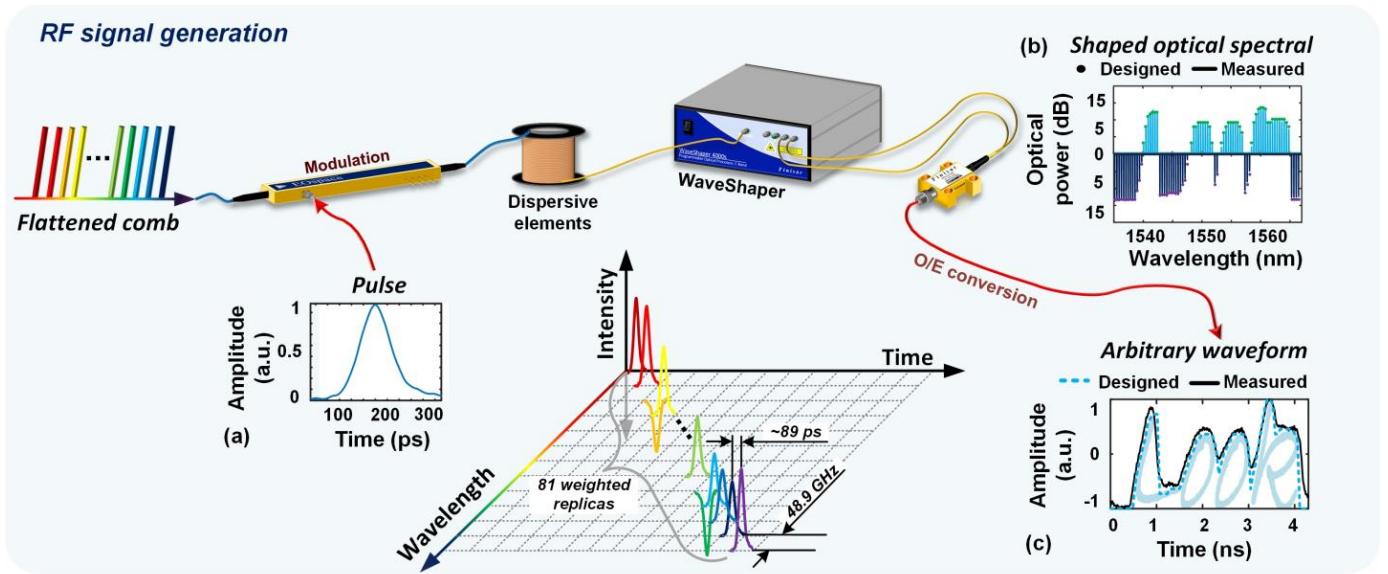

Fig. 3.　Conceptual diagram of arbitrary RF waveform generation.　(a) Temporal intensity waveform of the RF pulse. (b) Designed and measured optical spectral of the shaped micro-comb for arbitrary waveform generation. (c) Designed and measured temporal output intensity arbitrary waveform.

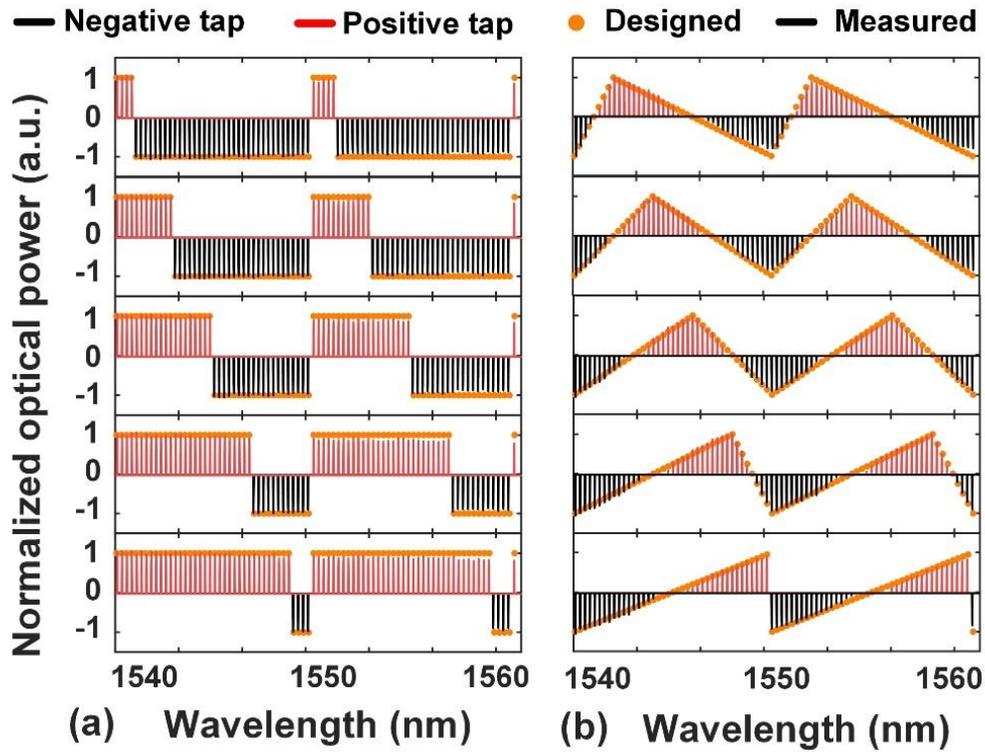

Fig. 4. Designed and measured optical spectra for (a) Square waveforms. (b) Sawtooth waveforms.

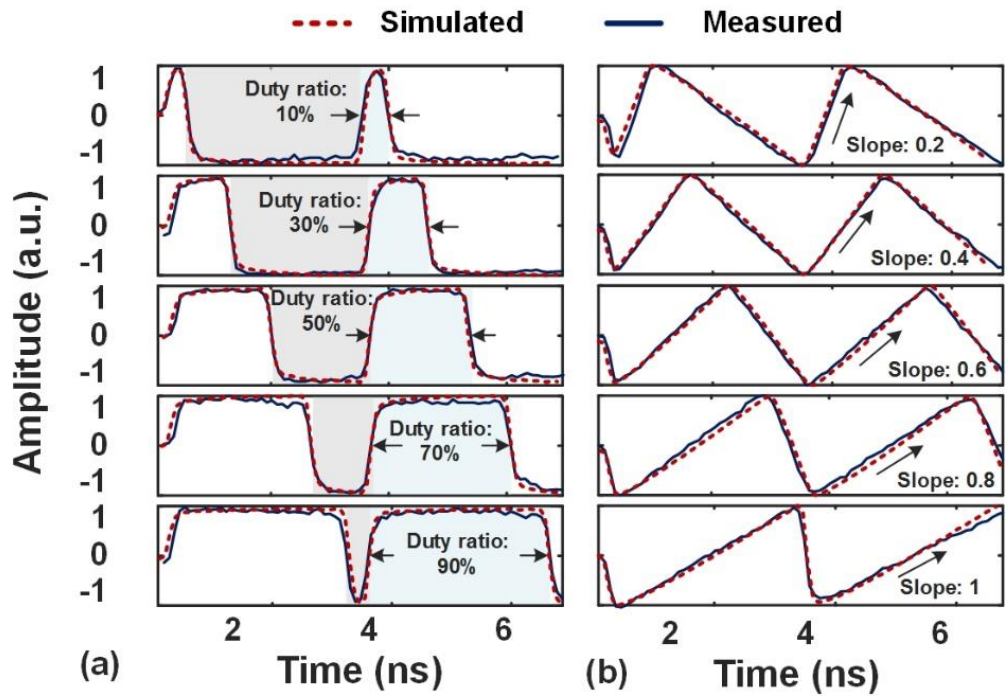

Fig. 5. Simulated and measured tunable RF waveforms. (a) Square waveforms. (b) Sawtooth waveforms.

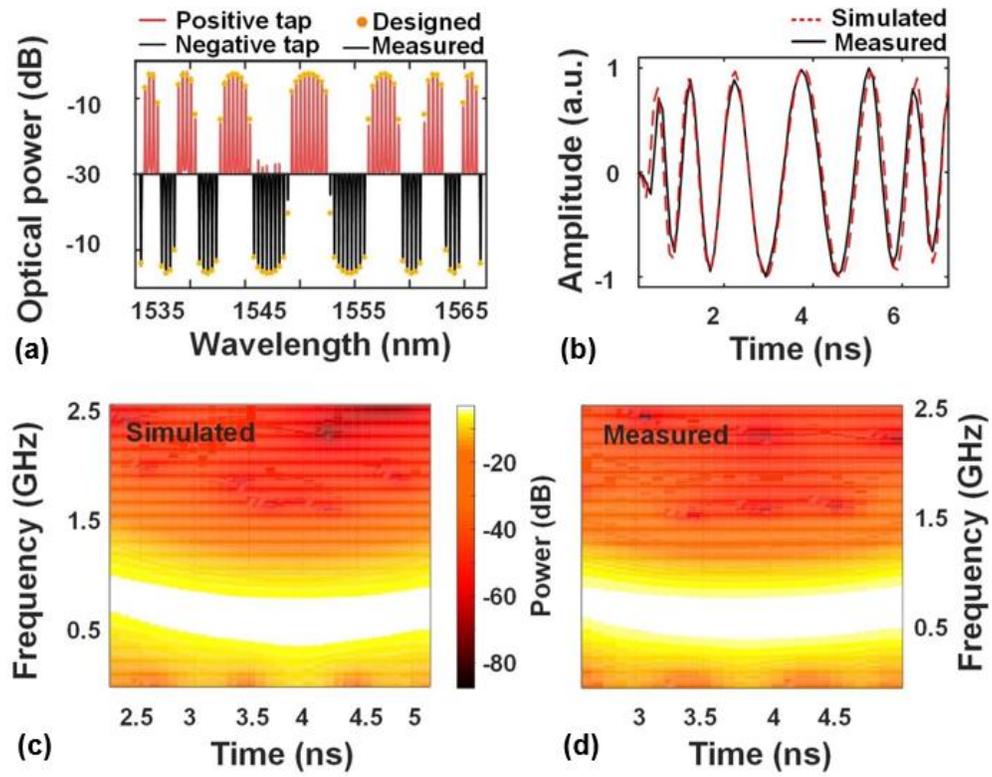

Fig. 6. Experimental results of the generated chirped waveform. (a) Designed and measured optical spectra. (b) Simulated and measured RF waveforms. The extracted corresponding instantaneous frequency (c) simulated (g) experimental results.